\documentclass[10pt, a4paper]{article}

\usepackage[final]{lrec2026}
\usepackage{color}
\usepackage{float}
\usepackage{amsmath}
\usepackage{amsfonts}
\usepackage{tabularray}
\usepackage[table]{xcolor}
\usepackage{booktabs,tabularx}
\usepackage{multirow}

\usepackage{comment}
\usepackage{graphicx}
\usepackage{subcaption}

\title{HARNESS: Lightweight Distilled Arabic Speech Foundation Models}

\name{Vrunda N. Sukhadia $^{1*}$, Shammur Absar Chowdhury$^2$}

\address{$^1$Amazon India, $^2$Qatar Computing Research Institute, HBKU, Qatar \\
         sukhadiavrunda@gmail.com, Shchowdhury@hbku.edu.qa\\}

\abstract{
Large self-supervised speech (SSL) models achieve strong downstream performance, but their size limits deployment in resource-constrained settings. We present HArnESS, an Arabic-centric self-supervised speech model family trained from scratch with iterative self-distillation, together with lightweight student variants that offer strong accuracy-efficiency trade-offs on Automatic Speech Recognition (ASR), Dialect Identification (DID), and Speech Emotion Recognition (SER). Our approach begins with a large bilingual Arabic-English teacher and progressively distills its knowledge into compressed student models while preserving Arabic-relevant acoustic and paralinguistic representations. We further study PCA-based compression of the teacher supervision signal to better match the capacity of shallow and thin students. Compared with HuBERT and XLS-R, HArnESS consistently improves performance on Arabic downstream tasks, while the compressed models remain competitive under substantial structural reduction. These results position HArnESS as a practical and accessible Arabic-centric SSL foundation for real-world speech applications.
 \\ }

\begin{document}

\maketitleabstract
\def\thefootnote{*}\footnotetext{This work was carried out at QCRI.}\def\thefootnote{\arabic{footnote}}

\section{Introduction}
\label{sec:introduction}
Self-supervised learning (SSL) has transformed speech processing by learning transferable representations from large amounts of unlabeled audio. Large SSL models capture rich acoustic and linguistic structure and have shown strong performance across a wide range of speech tasks \cite{chen2022wavlm, hsu2021hubert,baevski2022data2vec,mohamed2022self,chung2021w2v,yang2021superbspeechprocessinguniversal}. These models can be used either as fixed feature extractors or fine-tuned with limited labeled data, making them especially attractive in low-resource settings.

The effectiveness of SSL models, however, depends heavily on the scale, diversity, and balance of the pretraining data. Multilingual SSL models such as XLS-R \cite{babu2021xlsr} have shown clear advantages for low-resource languages compared with monolingual models trained on high-resource languages such as English \cite{shi2023mlsuperbmultilingualspeechuniversal}. At the same time, recent evidence suggests that multilingual models may disproportionately favor languages with greater pretraining coverage, which can limit gains for underrepresented languages \cite{storey2024language}. This motivates a closer study of language-focused SSL models that better reflect the linguistic and acoustic properties of a target language.

Arabic is a particularly challenging case for speech modeling. It is spoken across 22 countries and exhibits substantial dialectal diversity, with wide varieties differing in phonetics, morphology, and lexical usage. In addition, Arabic speech often includes influences from other languages, including English and French \cite{ali_connecting_2021}. This diversity makes Arabic speech processing difficult for generic multilingual models, which may not fully capture dialect-sensitive and culturally grounded speech patterns. These challenges motivate Arabic-centric SSL modeling that can better represent spoken Arabic while remaining robust to variation across regions and speaking styles.

At the same time, training and deploying language-focused SSL models remains expensive. Large-scale pretraining requires substantial compute, long training times, and broad unlabeled speech collections. These costs also make deployment difficult in practical and resource-constrained environments, where model size, memory use, and latency matter. Model compression is therefore essential for making such systems more accessible and usable.

Knowledge distillation has emerged as an effective approach for compressing large speech models while preserving much of their performance. In this setting, a smaller student model learns from a larger teacher model, leading to lower memory usage and faster inference with limited degradation in downstream quality. Prior work, including DistillHuBERT \cite{chang2022distilhubert}, FitHuBERT \cite{lee2022fithubert}, DPHuBERT \cite{peng2023dphubert}, SKILL \cite{zampierin2024skill}, and related methods \cite{ashihara2022deep,wang2022lighthubert}, has explored task-agnostic distillation for HuBERT-style models. However, such work has focused primarily on general-purpose compression, with limited attention to Arabic-centric SSL trained from scratch and systematically distilled into lightweight models.

While prior studies have applied self-supervised speech models such as HuBERT and wav2vec~2.0 to Arabic, Arabic-centric SSL remains under-explored, especially in the setting of large-scale training from scratch and deployment-oriented compression. In this work, we focus on both aspects. We introduce \textbf{H}uBERT-based \textbf{Ar}abic a\textbf{n}d \textbf{E}nglish \textbf{S}elf-\textbf{S}upervised Speech (\textbf{HArnESS}), an Arabic-centric SSL model family trained from scratch on large-scale bilingual Arabic-English speech, and we study iterative self-distillation to build compact student models that retain strong performance on ASR, SER, and DID.

We adopt bilingual Arabic-English pretraining for two reasons. First, English corpora provide additional acoustic and phonetic diversity at scale, which can stabilize representation learning when Arabic resources are comparatively limited and heterogeneous. Second, Arabic speech in real-world settings often includes borrowed English words and code-switching, particularly in conversational and media domains. Our goal is therefore not to weaken Arabic-centric modeling, but to combine Arabic-focused coverage with the regularization benefits of broader bilingual pretraining.

Following the HuBERT training paradigm, we first train a large teacher model, HArnESS-L, with 24 encoder layers through iterative self-distillation. We then transfer its knowledge to smaller students, yielding HArnESS-S, a shallow variant, and HArnESS-ST, a shallow (S) and thin (T) variant. In addition, we investigate low-rank approximation of the teacher supervision signal to simplify the distillation target space and improve knowledge transfer to compact students.

We evaluate HArnESS-L, HArnESS-S, and HArnESS-ST on three downstream tasks spanning content, dialectal, and paralinguistic information, namely ASR, DID, and SER. We compare them against HuBERT-Large, trained primarily on English, and XLS-R, a multilingual SSL model \cite{babu2021xlsr}. Our results show that Arabic-centric pretraining combined with iterative distillation provides an effective balance between task performance and model efficiency.

\begin{figure*}[!ht]
    \begin{center}   
    \includegraphics[scale=0.5]{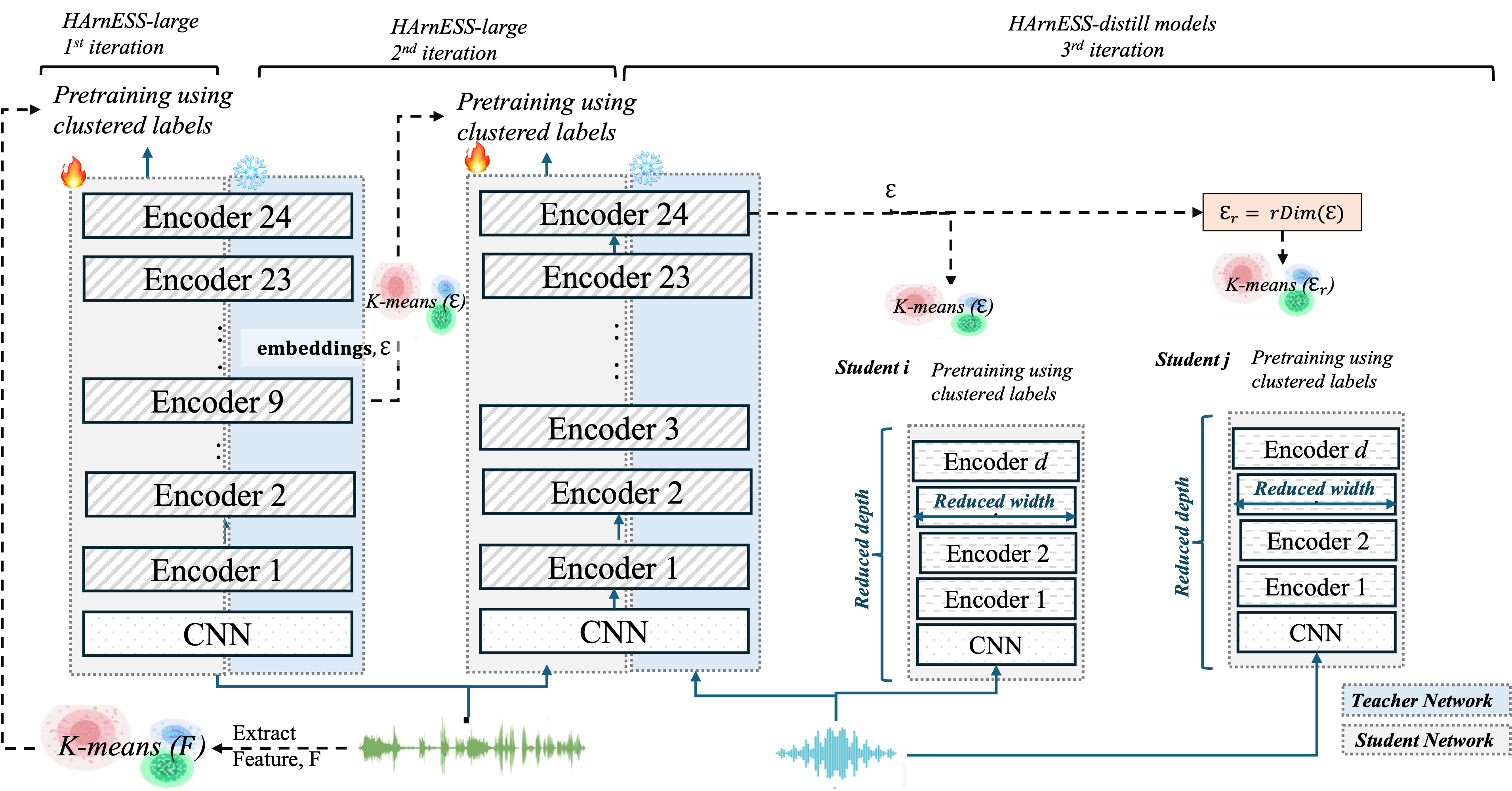}        
    \caption{Overview of the iterative self-distillation framework used to build the HArnESS model family.}
    \label{fig:overview}
    \end{center}
\end{figure*}

\noindent Our main contributions are as follows:
\begin{enumerate}
    \item We introduce HArnESS, an Arabic-centric SSL model family trained from scratch, consisting of HArnESS-L (large), HArnESS-S (shallow), and HArnESS-ST (shallow and thin).
    \item We study iterative self-distillation as a strategy for compressing Arabic-centric SSL models into lightweight deployment-oriented students.
    \item We investigate compact supervision through low-rank approximation of the teacher signal and analyze its effect on student performance.
    \item We benchmark the HArnESS family on ASR, DID, and SER, covering content, speaker-related, and paralinguistic speech tasks.
    \item We publicly release the distilled models and benchmark resources to support future research.\footnote{\url{https://huggingface.co/QCRI/distillHarness}}
\end{enumerate}

\section{HArnESS Models}
\label{sec:Harness_model}
\subsection{HArnESS Model}
\label{sec:HArnESS_model}

Figure~\ref{fig:overview} illustrates the HArnESS training pipeline, which follows a HuBERT-style iterative self-distillation procedure. At iteration $i$, we train a model $\mathbf{M}_i$ using discrete pseudo-labels produced from the previous iteration model $\mathbf{M}_{i-1}$. The core idea is masked-prediction pretraining: a subset of time frames is masked, and the model is optimized to predict the corresponding pseudo-labels.

\paragraph{Iterative self-distillation pipeline.}
Given an input utterance $x$, we first obtain frame-level embeddings from $\mathbf{M}_{i-1}$ and convert them into discrete targets via clustering (described below), yielding a pseudo-label sequence
$z^{(i-1)} = \{ z^{(i-1)}_t \}_{t=1}^{T}$ with $z^{(i-1)}_t \in \{1,\dots,K\}$.
We then train $\mathbf{M}_i$ to predict these targets from contextualized frame representations, where some frames are replaced by a mask token (or masked spans), forcing the model to use broader context.

\paragraph{Training regimen and compression schedule.}
We perform multiple iterations of refinement. In the first two iterations, we keep the model architecture unchanged to encourage progressively stronger acoustic abstractions. Starting from the third iteration, we distill and compress the model to obtain efficient variants. Specifically, we explore three compression axes:
\emph{(a)} reducing Transformer depth ($d$) to obtain a shallower model;
\emph{(b)} reducing model width (encoder dimension, $enc_d$) to obtain a thinner model; and
\emph{(c)} reducing attention capacity by decreasing the number of attention heads ($attn$).
This schedule yields a teacher-like encoder in early iterations and a family of compact students in later iterations.

\paragraph{Model architecture.}
HArnESS consists of a convolutional (CNN) feature extractor followed by a stack of Transformer encoder layers.
Similar to HuBERT encoders, the CNN front-end comprises 7 temporal convolution layers that transform raw audio into latent frame features.
The Transformer encoder contains $d$ layers with hidden dimension, ${\text{emb}_d}$. 
Each layer includes multi-head self-attention (MHA) with $attn$ heads and a position-wise feed-forward network (FFN).
A linear prediction head maps contextualized frame representations to a categorical distribution over $K$ cluster IDs.


\paragraph{Training objective.}
HArnESS is trained with a HuBERT-style masked prediction objective. For each utterance, we mask a subset of time frames (or spans) in the input representation and train $\mathbf{M}_i$ to predict the corresponding discrete pseudo-labels generated from $\mathbf{M}_{i-1}$. We use standard cross-entropy classification over the cluster IDs. To improve stability, we compute the loss on both masked and unmasked frames and combine them with a fixed weighting: the masked-frame loss encourages contextual reasoning over surrounding speech, while the unmasked-frame loss provides an additional learning signal that helps prevent training collapse and improves convergence.

\paragraph{Pseudo-label generation.}
To obtain discrete targets for iteration $i$, we extract frame-level embeddings from the previous model $\mathbf{M}_{i-1}$ and cluster them using $K$-means, producing a pseudo-label sequence
$z^{(i-1)}=\{z^{(i-1)}_t\}_{t=1}^{T}$ with $z^{(i-1)}_t \in \{1,\dots,K\}$.
Unless otherwise stated, we use last-layer embeddings, which provide the most abstract and stable representations.\footnote{We also explored averaged embeddings from selected layers and observed no consistent gains.}

\paragraph{PCA for supervision signal compression.}
We also investigate PCA as a way to compress the teacher supervision signal before clustering. Instead of clustering the full teacher embedding, we optionally project it to a lower-dimensional space and then derive pseudo-labels from the projected representation. 
Concretely, given an embedding vector $h_t \in \mathbb{R}^{D}$ from $\mathbf{M}_{i-1}$, we apply PCA to obtain a compressed representation $\tilde{h}_t \in \mathbb{R}^{D'}$ ($D' \ll D$) and then cluster $\tilde{h}_t$.

This serves two purposes. First, dimensionality reduction can remove noisy or redundant directions, which may improve clustering robustness. Second, it produces a simpler target space that better matches the capacity of compressed students, especially when model width ($enc_d$) is reduced. In this sense, PCA does not compress the student input directly; rather, it simplifies the discrete targets used during distillation.

\paragraph{Initialization and iteration-specific supervision.}
For the initial iteration $i=1$, we bootstrap pseudo-labels by extracting MFCC features from raw speech $x$ and clustering them to obtain $z^{(0)}$. For the next iteration ($i=2$), we generate pseudo-labels from intermediate representations of $\mathbf{M}_{0}$, using the 9\textsuperscript{th} Transformer layer embeddings for clustering. For all subsequent iterations ($i \ge 3$), we generate pseudo-labels using last-layer embeddings from $\mathbf{M}_{i-1}$, which provide the most abstract and stable representations.

For training $\mathbf{M}_i$ ($i \ge 1$), we explore two weight initialization strategies:
\emph{(a)} random initialization (uniformly sampled weights), and
\emph{(b)} blocked-averaging initialization, where groups of student layers are initialized by averaging corresponding blocks of layers from $\mathbf{M}_{i-1}$.
Blocked averaging provides a smoother transition across iterations and often improves stability, particularly when compressing depth/width in later iterations.

\section{Experimental Setups}
\label{sec:Experimental_setups}

\subsection{Pre-training Data}
\label{sec:pretrain_data}

\noindent\textbf{Iterations 1--2 (bilingual pretraining).}
We pre-train HArnESS on a mixture of publicly available Arabic and English speech corpora, including QASR~\citeplanguageresource{mubarak_qasr_2021}, MGB3~\citeplanguageresource{ali2017speech}, LibriSpeech~\citeplanguageresource{panayotov2015librispeech}, Common Voice (Arabic/English)~\citeplanguageresource{ardila2020commonvoicemassivelymultilingualspeech}, and GigaSpeech~\citeplanguageresource{chen2021gigaspeechevolvingmultidomainasr}, among others. The base pretraining pool consists of approximately \textbf{4K hours of Arabic} and \textbf{3.56K hours of English} speech. We further expand the training data through augmentation to reach approximately \textbf{23K hours} in total. Of this augmented portion, around \textbf{300 hours} come from additive background-noise augmentation, while the remainder is primarily produced through SpecAugment-based transformations.

To improve dialectal and cultural coverage, we also incorporate spoken content from 15 Arabic-speaking countries crawled from YouTube, covering diverse Arabic dialects. We provide a coarse dialect breakdown of the Arabic data by major region in Table~\ref{tab:combined_training_data}, grouping samples into MSA, Gulf, Levantine, Egyptian, Maghrebi, mixed, and unlabeled categories where exact dialect labels are unavailable. All official development and test partitions are excluded from pretraining to avoid data leakage.


\begin{table}[ht]
\centering

\scalebox{0.63}{
\renewcommand{\arraystretch}{1.1}
\begin{tabular}{@{}lcr@{}}
\toprule
\textbf{Category} & \textbf{Sub-category / Dialect} & \textbf{Duration (Hrs)} \\ \midrule
\textbf{Original Clean Data} & & \textbf{7,566.00} \\
& English Subset & 3,565.00 \\
& Arabic Subset & 4,001.00 \\
\cmidrule(lr){2-3}
& \hspace{3mm} MSA / General Arabic & 3,603.28 \\
& \hspace{3mm} Levantine & 107.69 \\
& \hspace{3mm} Egyptian & 109.20 \\
& \hspace{3mm} Gulf & 77.13 \\
& \hspace{3mm} Maghrebi & 69.11 \\
& \hspace{3mm} Other & 34.59 \\ \midrule
\textbf{Augmented Data} & & \textbf{15,434.00} \\
& Speed Perturbation ($0.9\times, 1.1\times$) & 15,134.00 \\
& Noise Augmentation (Arabic) & 300.00 \\ \midrule
\textbf{Total Training Volume} & & \textbf{23,000.00} \\ \bottomrule
\end{tabular}}
\caption{Comprehensive breakdown of the 23,000-hour training corpus, including language distribution, dialectal variety, and augmentation strategies.}
\label{tab:combined_training_data}
\end{table}

\noindent\textbf{Iteration 3 (Arabic-only distillation).}
Our primary goal is to obtain lightweight Arabic-centric models.
Accordingly, for the distillation/compression iteration, we use approximately \textbf{1{,}100 hours} of Arabic speech drawn from the QASR training data.
For $K$-means training in this phase, we randomly sample \textbf{30\%} of the iteration-3 data (approximately \textbf{300 hours}) to reduce clustering cost while maintaining linguistic diversity.

\subsection{Downstream Tasks and Data}
\label{sec:downstream_data}

Benchmarking SSL speech encoders for English is supported by standardized suites such as SUPERB~\cite{yang2021superbspeechprocessinguniversal}. In contrast, Arabic speech lacks an analogous standardized benchmark.
To address this gap, we evaluate HArnESS across three representative Arabic tasks:
\textbf{ASR} (content recognition),
\textbf{dialect identification (DID)} (speaker information), and
\textbf{speaker emotion recognition (SER)} (paralinguistic analysis).

\noindent\textbf{ASR.}
We fine-tune on a \textbf{300-hour} subset of QASR and evaluate on the MGB2 \cite{ali2019mgb2challengearabicmultidialect} test set. To assess out-of-domain generalization, we additionally report performance on the MGB3 test set.

\noindent\textbf{SER.}
We use KSUEmotion~\citeplanguageresource{9393909}, collected from 23 speakers with six emotion classes. The dataset is split into train (3.30 h), dev (0.83 h), and test (1.0 h).\footnote{We will release our split for reproducibility.}

\noindent\textbf{DID.}
We use the ADI5 dataset with five region-based dialect classes (MSA, Egyptian, Levantine, North African, and Gulf) and the official train/dev/test splits.

\noindent\textbf{Metrics.}
We report word error rate (WER) for ASR and classification accuracy (Acc) for DID and SER.

\begin{table*}[t]
\centering
\scalebox{0.8}{
\begin{tabular}{l|c}
\toprule

\midrule
Total pre-training audio (Iter 1--2) & 23k hours (Arabic/English $\approx$ balanced) \\
$K$-means training subset (Iter 1--2) & 300 hours \\
Distillation audio (Iter 3) & 1{,}100 hours (Arabic-only) \\
$K$-means training subset (Iter 3) & 30\% $\approx$ 300 hours \\
\midrule
\# clusters ($K$) & 1000 \\
Feature type for $i{=}0$ targets & MFCC (39-dim) \\
Embedding layer for $i{=}1$ targets & Layer 9 embeddings of $\mathbf{M}_0$ \\
Embedding layer for $i{\ge}2$ targets & Last-layer embeddings of $\mathbf{M}_{i-1}$ \\
\midrule
Iteration 1 steps / GPUs / batch & 500k / 24$\times$H100 / 62.5s audio per GPU \\
Iteration 2 steps / GPUs / batch & 700k / 24$\times$H100 / 62.5s audio per GPU \\
Iteration 3 steps / GPUs / batch & 300k / 8$\times$H100 / 75s audio per GPU \\
Mask probability ($p_{\text{mask}}$) & 0.80 \\
Mask span length (frames) & 10 \\ \midrule
PCA dimension ($D'$; when enabled) & 512 \\
\bottomrule
\end{tabular}}
\caption{Key pre-training hyperparameters.}
\label{tab:pretrain_hparams}
\end{table*}

\subsection{Pre-training Hyperparameters}
\label{sec:pretrain_hparams}

We train HArnESS using the \texttt{fairseq} codebase~\cite{ott2019fairseq}. Table~\ref{tab:pretrain_hparams} summarizes the key hyperparameters. 
Unless stated otherwise, we only change the supervision source and model capacity (Table~\ref{tab:arch_desc}).

\subsection{Pre-training Procedure}
\label{sec:pretrain_params}
\begin{table*}[!ht]
\centering

\scalebox{0.8}{
\begin{tabular}{l|cccccc}
\toprule
\textit{Models} & \textbf{XR} & \textbf{HuL} & \textbf{H-L} & \textbf{H-S} & \textbf{H-ST} & \textbf{H-ST (PCA)} \\  
\midrule
\textit{Supervision} 
& -- & -- 
& \shortstack{$L9_{\text{emb}}$\\($i=1$)} 
& \multicolumn{2}{c}{\shortstack{$L23_{\text{emb}}$\\($i=2$)}} 
& \shortstack{PCA($L23_{\text{emb}}$)\\($i=2$)} \\  
\midrule
\multicolumn{7}{c}{\textbf{CNN Encoder}} \\  
\midrule
Strides       & \multicolumn{6}{c}{5, 2, 2, 2, 2, 2, 2} \\  
Kernel Width  & \multicolumn{6}{c}{10, 3, 3, 3, 3, 2, 2} \\  
Channels      & \multicolumn{6}{c}{512} \\  
\midrule
\multicolumn{7}{c}{\textbf{Transformer}} \\  
\midrule
Depth ($l$)              & 24 & 24 & 24 & 4 & 4 & 4  \\  
Emb. Dim (${\text{emb}_d}$) & 1024 & 1024 & 1024 & 1024 & 512 & 512 \\  
FFN Dim ($d_{\text{ffn}}$)  & 4096 & 4096 & 4096 & 2048 & 2048 & 2048 \\  
Attn. Heads ($h_{\text{attn}}$) & 16 & 16 & 16 & 16 & 16 & 16 \\  
\midrule
\multicolumn{7}{c}{\textbf{Projection}} \\  
\midrule
Dim. ($d_p$) & 768 & 768 & 768 & 768 & 768 & 768 \\  
\midrule
\multicolumn{7}{c}{\textbf{Params}} \\  
\midrule
\textit{in M} & 300 & 316 & 316 & 65 & 28 & 28 \\  
\bottomrule
\end{tabular}}
\caption{SSL Model Architecture Comparison. XR: XLS-R, HuL: HuBERT-Large, H-L: HArnESS-Large, H-S: HArnESS-Shallow, H-ST: HArnESS-Shallow and Thin. Dim. dimension, Emb.: Embedding. $L*_{emb}$: Embedding from layer * (e.g. 23) of model from iteration $i$. }
\label{tab:arch_desc}
\end{table*}

Model configurations for the upstream encoders are summarized in Table~\ref{tab:arch_desc}.

\paragraph{Iterations 1--2 (HArnESS-L).}
For the first two iterations, we train the large model (\textbf{HArnESS-L}; 24 Transformer layers) on the 23k-hour bilingual mixture.
Iteration 1 is trained for 500k steps and iteration 2 for 700k steps.
For iteration 1 pseudo-labels, we cluster 39-dimensional MFCC features with $K{=}1000$ clusters.
For iteration 2 pseudo-labels, we extract latent representations from the \textbf{9th Transformer layer} of the iteration-1 model and cluster them with $K{=}1000$ clusters to obtain refined targets.

\paragraph{Iteration 3 (compressed students).}
For iteration $i{=}3$, we train compressed models using \textbf{HArnESS-S} and \textbf{HArnESS-ST}, both with a 4-layer Transformer encoder and reduced capacity (Table~\ref{tab:arch_desc}).
Pseudo-labels are generated by clustering \textbf{last-layer} embeddings from the iteration-2 HArnESS-L teacher with $K{=}1000$ clusters.
Thus, HArnESS-S and HArnESS-ST correspond to the third-iteration distilled models trained on 1{,}100 hours of Arabic speech.
\vspace{-2mm}
\subsection{Downstream Training}
\label{sec:downstream_train}
For downstream evaluation, we keep the SSL encoder frozen and use it strictly as a feature extractor. We obtain frame-level representations from all Transformer layers, average them to form an utterance-level representation, and train task-specific downstream models on top of these fixed features only. No gradients are propagated into the SSL encoder during downstream training.

\subsubsection{DID and SER Architecture}
\label{sec:did_ser_arch}

For DID and SER, we train a lightweight classifier on top of the frozen SSL features with a batch size of 4 for 10k steps. The classifier consists of three temporal convolution layers with kernel size 5, ReLU activations, and dropout of 0.4, followed by self-attention pooling, a feed-forward layer, and a final softmax layer. All hidden dimensions are set to 80. This setup isolates the quality of the learned SSL representations by keeping the encoder fixed and limiting the trainable parameters to the downstream classifier.


\subsubsection{ASR Architecture}
\label{sec:asr_arch}

For ASR, we train an encoder--decoder model with a joint CTC/attention objective using the ESPnet toolkit.\footnote{Using ESPnet toolkit.}
The encoder consists of two Conformer layers and the decoder consists of two Transformer layers, each with 8 attention heads and 2048 linear units.
We train for 70 epochs.

\section{Results}
\label{sec:results}
\begin{figure}[!h] 
    \centering
    \begin{subfigure}[b]{0.45\textwidth}
        \centering
        \includegraphics[width=\textwidth]{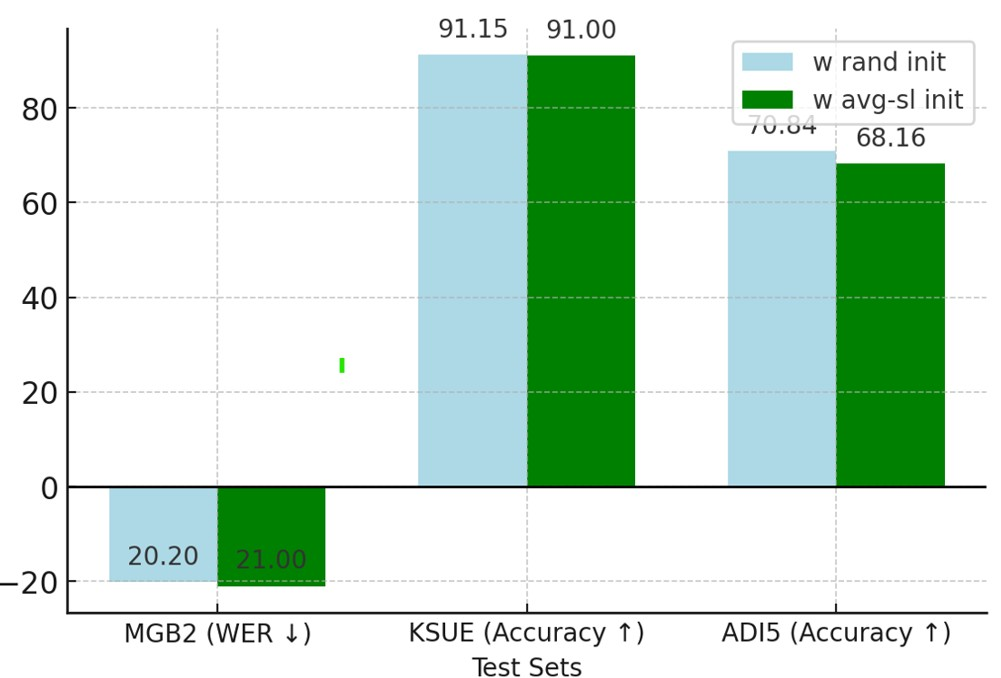}
        \caption{Weight Initialization}
        \label{fig:ablation_1}
    \end{subfigure}
    \hfill 
    \begin{subfigure}[b]{0.45\textwidth}
        \centering
        \includegraphics[width=\textwidth]{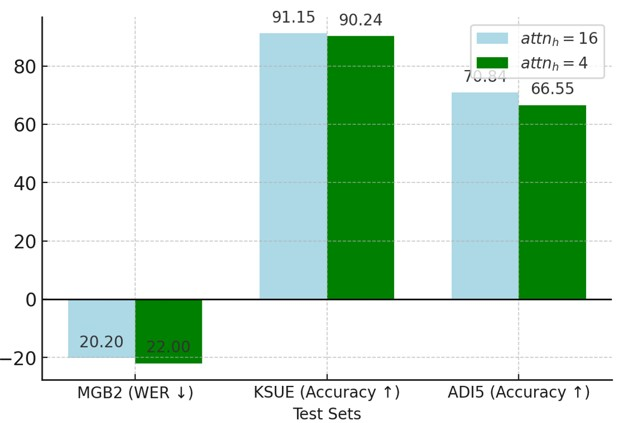}
        \caption{Number of attention heads}
        \label{fig:ablation_2}
    \end{subfigure}
    \hfill

    \begin{subfigure}[b]{0.5\textwidth}
        \centering
        \includegraphics[width=\textwidth]{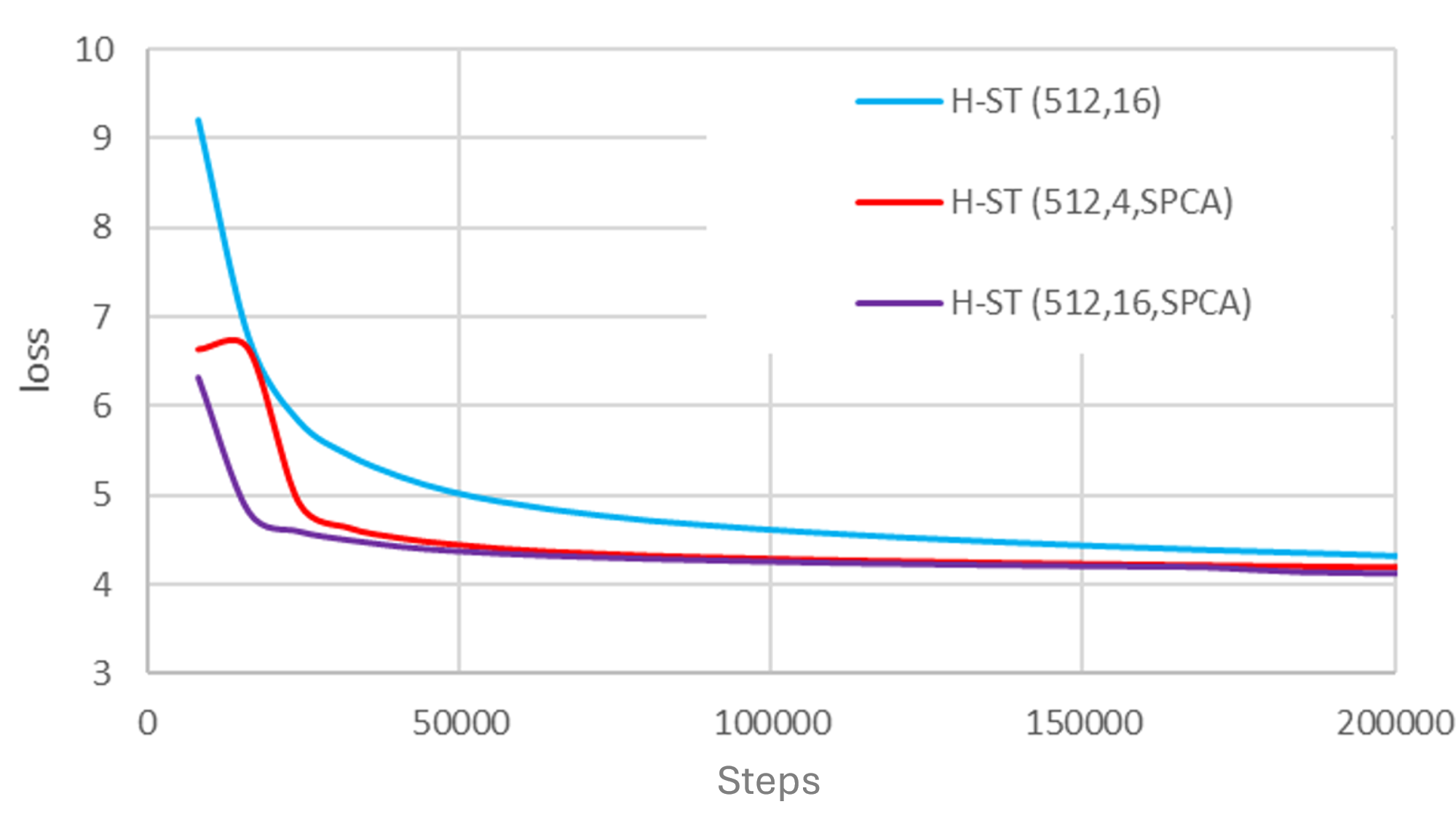}
        \caption{Effect of dimension reduction} 
        \label{fig:ablation_3}
    \end{subfigure}

\caption{Ablation results for the compressed student models. \subref{fig:ablation_1}) Effect of student initialization strategy. (\subref{fig:ablation_2}) Effect of reducing the number of attention heads. (\subref{fig:ablation_3}) Effect of applying PCA to teacher embeddings before clustering when generating pseudo-labels for student training. H-ST($emb_d$, $attn_h$), SPA means PCA applied for supervision.}
    \label{fig:abl}
\end{figure}

\begin{table*}[h]
\centering
\scalebox{0.7}{
\begin{tabular}{l|cc|c|c}
\toprule
\multirow{2}{*}{\textbf{Models}} & \multicolumn{2}{c|}{\textbf{ASR (WER $\downarrow$)}} & \textbf{SER (Acc $\uparrow$)} & \textbf{DID (Acc $\uparrow$)} \\
\cmidrule(lr){2-3} \cmidrule(lr){4-4} \cmidrule(lr){5-5}
 & \textbf{MGB2} & \textbf{MGB3} & \textbf{KSUEmotion} & \textbf{ADI5} \\
\midrule

\multicolumn{5}{c}{\textit{Published task-specific reference results (context only)}} \\
\midrule
Reference result & 
\begin{tabular}{@{}c@{}} 10.24 \\ \cite{fanarteam2025} \end{tabular} & 
\begin{tabular}{@{}c@{}} 21.31 \\ \cite{fanarteam2025} \end{tabular} & 
\begin{tabular}{@{}c@{}} 85.53\% \\ \cite{abouzeid2025arabemonetlightweighthybrid2d} \end{tabular} & 
\begin{tabular}{@{}c@{}} 82.5\% \\ \cite{kulkarni-aldarmaki-2023-yet} \end{tabular} \\
\midrule
\midrule\midrule

\multicolumn{5}{c}{\textit{Our downstream evaluation with frozen SSL encoders}} \\
\midrule
\textbf{HuBERT-L} (\textit{English}) & 22.6$^*$ & 51.2$^*$ & 91.92\% & 64.14\% \\
\textbf{XLS-R} (\textit{Multilingual}) & 22.60$^*$ & 51.80$^*$ & 73.32\% & 42.35\% \\
\textbf{HArnESS-L} (\textit{Bilingual: Arabic-English}) & \textbf{15.50}$^*$ & \textbf{41.60}$^*$ & \textbf{94.66\%} & \textbf{84.98\%} \\

\midrule\midrule

\multicolumn{5}{c}{\textit{Compressed HArnESS students distilled with $\approx$1000h Arabic-only data}} \\
\midrule
\textbf{HArnESS-S} ($\Delta S=79.4\%$) & \textit{20.20}$^*$ & \textit{52.80}$^*$ & \textit{91.15\%} & \textit{70.84\%} \\
\textbf{HArnESS-ST} ($\Delta S=93.7\%$) & 23.20$^*$ & 58.20$^*$ & 89.02\% & 69.77\% \\
\textbf{HArnESS-ST$^\Xi$} ($\Delta S=93.7\%$) & 22.50$^*$ & 55.60$^*$ & 87.34\% & 61.64\% \\
\bottomrule
\end{tabular}}
\caption{Performance comparison on ASR, SER, and DID. ASR downstream results in our setup are obtained by training on a \textbf{ 300h QASR} subset, results denoted by a $^*$. The top block lists previously published task-specific reference results from separate studies and is included only for context. These numbers are not from a unified baseline and are not directly comparable to the models evaluated in our setup. L denotes the large teacher model, S the shallow student, and ST the shallow-thin student. $\Delta S$ denotes overall structural compression relative to HArnESS-L.}
\label{tab.rslt}
\end{table*}

 \paragraph*{Comparison with Upper Bound: SOTA Model} 
 As contextual reference, Table~\ref{tab.rslt} also reports representative published results from strong task-specific systems for Arabic ASR, DID, and SER. These results come from separate studies and are not directly comparable to our models because they differ in architecture, supervision, training data, and experimental protocol. We therefore use them only to contextualize performance, not to claim direct state-of-the-art results.
 For ASR, we report results alongside Fanar ASR \cite{fanarteam2025}, a specialized system trained on more than 10K hours of MSA and dialectal Arabic speech. Under our much more constrained setup, where ASR fine-tuning uses only 300 hours of MSA data, HArnESS-L remains within about 5 WER points on MGB2 and MGB3, and HArnESS-S within about 10 points. For DID and SER, HArnESS-L also shows strong results relative to the published reference numbers, achieving 84.98\% on ADI5 compared with 82.5\% \cite{kulkarni-aldarmaki-2023-yet}, and 94.66\% on KSUEmotion compared with 85.53\% (on ResNet-based architecture). Although these comparisons are only approximate, they indicate that HArnESS-L is competitive with strong specialized systems, while the distilled students preserve much of the teacher's performance with substantially smaller capacity.

\paragraph*{HArnESS-L {\em  vs.} Existing SSLs for Arabic}

Compared with HuBERT-L and XLS-R, HArnESS-L performs better across the evaluated Arabic tasks, suggesting that Arabic-centric pretraining is beneficial for downstream Arabic speech processing. The compressed HArnESS variants also outperform the multilingual XLS-R baseline on several tasks, indicating that iterative distillation preserves useful task-relevant structure even under heavy compression.

\begin{table}[h]
\centering

\scalebox{0.75}{

\begin{tabular}{l|ccc}
\toprule
\textbf{Test Sets} & \textbf{$emb_d$=1024} & \textbf{$emb_d$=512} & \textbf{$emb_d$=256} \\ 
\midrule
MGB2 (WER ↓)  & 20.2  & 23.20    & 22.3  \\  
KSUEmotion (Acc ↑)  & 91.15\%  & 89.02\%  & 79.42\%  \\  
ADI5 (Acc ↑)  & 70.84\%  & 69.77\%  & 53.41\%  \\  
\midrule
\textbf{$\Delta S$} & 70.43\%  & 91.14\%  & 96.52\%  \\
\bottomrule

\end{tabular}}
\caption{Performance Comparison for different embedding dimensions. $\Delta S$: Overall structural compression.}
\label{tab.embred}
\end{table}
\paragraph*{Effects of structural compression and design choices.}
Figure~\ref{fig:abl} presents three ablation studies on student design, covering initialization, structural compression, and supervision compression.
For iteration $i=3$, we first examined the effect of student weight initialization and observed only minor differences in downstream performance (Figure~\ref{fig:abl}). This indicates that initialization plays a limited role at this stage, and that performance depends more strongly on the distilled supervision signal.

We then evaluated the effect of reducing model depth. HArnESS-S achieves 79.4\% structural compression relative to HArnESS-L while maintaining strong performance across tasks. Despite this compression, it still outperforms the multilingual and English SSL baselines, highlighting the effectiveness of Arabic-centric distillation. Compared with HArnESS-L, however, HArnESS-S shows a 4.7 absolute increase in WER, a 3.51-point drop in SER accuracy, and a 14.4-point drop in DID accuracy. The larger degradation on DID suggests that dialect-related cues are harder to preserve in shallower architectures.

Next, we reduced the number of attention heads from HArnESS-S ($\textit{attn}=16$) to HArnESS-S$^*$ ($\textit{attn}=4$), which yields an additional 26.15\% structural compression, reducing the model from 65M to 48M parameters. This change has only a limited effect on ASR and SER, but causes a larger drop on DID (Figure~\ref{fig:abl}), again indicating that dialect-sensitive information is more susceptible to architectural compression.

Finally, we examined embedding-dimension reduction (Table~\ref{tab.embred}). At extreme compression ($\Delta S=96.52\%$ relative to HArnESS-L), performance drops sharply across tasks. This result suggests that overly aggressive dimensionality reduction weakens the representational capacity of the student and substantially limits downstream performance.

\paragraph*{Effect of compressing the supervision signal.}
We also study whether simplifying the teacher supervision signal improves student training. Specifically, in iteration $i=3$, we compare knowledge distillation with and without applying PCA to the teacher embeddings before clustering. As shown in Figure~\ref{fig:ablation_3}, supervision derived from PCA-reduced embeddings converges faster than supervision from the original embeddings. This suggests that reducing redundancy in the teacher feature space produces a cleaner supervision signal, leading to more stable and efficient optimization while preserving effective knowledge transfer.

\section{Conclusion}

In this work, we introduced HArnESS, an Arabic-centric self-supervised speech model family designed to better capture the diversity of Arabic dialectal speech. Using an iterative self-distillation framework, we transferred knowledge from a large bilingual teacher model to compact shallow and shallow-thin student models while preserving Arabic-relevant speech representations. Experiments on Arabic ASR, SER, and DID show that HArnESS is competitive with, and in some cases stronger than, multilingual baselines such as HuBERT and XLS-R. The compressed HArnESS variants further offer an attractive efficiency-performance trade-off, making them suitable for more resource-constrained settings.
Our downstream evaluation relies on frozen encoders, providing a controlled assessment of representation quality, but it does not fully reflect the gains that may emerge under end-to-end fine-tuning. Future work will extend the comparison to fine-tuned settings and broader baselines. We will publicly release the lightweight models and benchmarking resources to facilitate future research.

\section{Bibliographical References}\label{sec:reference}

\bibliographystyle{lrec2026-natbib}
\bibliography{mybib}

\begin{thebibliography}{6}
\expandafter\ifx\csname natexlab\endcsname\relax\def\natexlab#1{#1}\fi

\bibitem[{Ali et~al.(2017)Ali, Vogel, and Renals}]{ali2017speech}
Ahmed Ali, Stephan Vogel, and Steve Renals. 2017.
\newblock Speech recognition challenge in the wild: {Arabic MGB-3}.
\newblock In \emph{2017 IEEE Automatic Speech Recognition and Understanding Workshop (ASRU)}, pages 316--322. IEEE.

\bibitem[{Ardila et~al.(2020)Ardila, Branson, Davis, Henretty, Kohler, Meyer, Morais, Saunders, Tyers, and Weber}]{ardila2020commonvoicemassivelymultilingualspeech}
Rosana Ardila, Megan Branson, Kelly Davis, Michael Henretty, Michael Kohler, Josh Meyer, Reuben Morais, Lindsay Saunders, Francis~M. Tyers, and Gregor Weber. 2020.
\newblock \href {http://arxiv.org/abs/1912.06670} {Common voice: A massively-multilingual speech corpus}.

\bibitem[{Chen et~al.(2021)Chen, Chai, Wang, Du, Zhang, Weng, Su, Povey, Trmal, Zhang, Jin, Khudanpur, Watanabe, Zhao, Zou, Li, Yao, Wang, Wang, You, and Yan}]{chen2021gigaspeechevolvingmultidomainasr}
Guoguo Chen, Shuzhou Chai, Guanbo Wang, Jiayu Du, Wei-Qiang Zhang, Chao Weng, Dan Su, Daniel Povey, Jan Trmal, Junbo Zhang, Mingjie Jin, Sanjeev Khudanpur, Shinji Watanabe, Shuaijiang Zhao, Wei Zou, Xiangang Li, Xuchen Yao, Yongqing Wang, Yujun Wang, Zhao You, and Zhiyong Yan. 2021.
\newblock \href {http://arxiv.org/abs/2106.06909} {Gigaspeech: An evolving, multi-domain asr corpus with 10,000 hours of transcribed audio}.

\bibitem[{Meftah et~al.(2021)Meftah, Qamhan, Seddiq, Alotaibi, and Selouani}]{9393909}
Ali~Hamid Meftah, Mustafa~A. Qamhan, Yasser Seddiq, Yousef~A. Alotaibi, and Sid~Ahmed Selouani. 2021.
\newblock \href {https://doi.org/10.1109/ACCESS.2021.3070751} {King saud university emotions corpus: Construction, analysis, evaluation, and comparison}.
\newblock \emph{IEEE Access}, 9:54201--54219.

\bibitem[{Mubarak et~al.(2021)Mubarak, Hussein, Chowdhury, and Ali}]{mubarak_qasr_2021}
Hamdy Mubarak, Amir Hussein, Shammur~Absar Chowdhury, and Ahmed Ali. 2021.
\newblock {QASR}: {QCRI} {Aljazeera} {Speech} {Resource}. {A} {Large} {Scale} {Annotated} {Arabic} {Speech} {Corpus}.
\newblock In \emph{{Proc. of the 59th Annual Meeting of the Association for Computational Linguistics (ACL)}}, pages 2274--2285, Online. Association for Computational Linguistics.

\bibitem[{Panayotov et~al.(2015)Panayotov, Chen, Povey, and Khudanpur}]{panayotov2015librispeech}
Vassil Panayotov, Guoguo Chen, Daniel Povey, and Sanjeev Khudanpur. 2015.
\newblock Librispeech: an asr corpus based on public domain audio books.
\newblock In \emph{2015 IEEE international conference on acoustics, speech and signal processing (ICASSP)}, pages 5206--5210.

\end{thebibliography}

\section{Language Resource References}
\label{lr:ref}
\bibliographystylelanguageresource{lrec2026-natbib}
\bibliographylanguageresource{languageresource}

\end{document}